\documentclass[12pt]{iopart}

\usepackage{graphicx}
\expandafter\let\csname equation*\endcsname\relax
\expandafter\let\csname endequation*\endcsname\relax
\usepackage{amsmath}
\usepackage{layouts}
\usepackage{hyperref}
\usepackage{comment}
\usepackage{float}

\begin{document}

%\printinunitsof{in}\prntlen{\linewidth}

\title[Electromagnetic Simulation of Heat Transport in Superconducting Qubits]{Electromagnetic Simulation and Microwave Circuit Approach of Heat Transport in Superconducting Qubits}
\author{Christoforus Dimas Satrya, Andrew Guthrie, Ilari Mäkinen and Jukka Pekola}
\address{Pico group, QTF Centre of Excellence, Department of Applied Physics, Aalto University School of Science, P.O. Box 13500, 00076 Aalto, Finland}
\ead{christoforus.satrya@aalto.fi, andrew.guthrie@aalto.fi, and ilari.makinen@aalto.fi}
\vspace{10pt}
\begin{indented}
\item[]%July 2022
\end{indented}

\begin{abstract}
The study of quantum heat transport in superconducting circuits is significant for further understanding the connection between quantum mechanics and thermodynamics, and for possible applications for quantum information. The first experimental realisations of devices demonstrating photonic heat transport mediated by a qubit have already been designed and measured. Motivated by the analysis of such experimental results, and for future experimental designs, we numerically evaluate the photonic heat transport of qubit-resonator devices in the linear circuit regime through electromagnetic simulations using Sonnet software, and compare with microwave circuit theory. We show that the method is a powerful tool to calculate heat transport and predict unwanted parasitic resonances and background.

%Furthermore, we discuss the simulation results in the context of the previous experimental results. 
\end{abstract}

\noindent{\it Quantum Thermodynamics, Superconducting Circuits, Superconductivity, Photonic Heat Transport, Quantum Information, Sonnet\/}

\section{Introduction}

%Superconducting circuits present a practical, controllable platform in which to realise quantum thermal devices \cite{RevModPhys.78.217, RevModPhys.93.041001}. Understanding the processes underpinning thermal transport in such mesoscopic structures has significant potential to further our understanding of quantum thermodynamics \cite{annurev-physchem-040513-103724,Alicki_1979,PhysRevE.76.031105, ,polini2022materials,Myers:2022lvm}. Advances in nanofabrication have allowed the study of heat transport in on-chip planar superconducting structures, and has been coined `circuits quantum thermodynamics', or cQTD \cite{Karimi2022}. Closely related to `circuit quantum electrodynamics' (cQED), cQTD describes the interplay between superconducting qubits and microwave cavities, however considers the inclusion of normal-metal elements providing sources of thermal photons in such circuits. 

Circuit quantum thermodynamics (cQTD) is an emerging field that studies thermodynamics of a quantum system interacting with dissipative environments, theorised and/or realised in the platform of superconducting and normal-metal circuits \cite{Karimi2022}. Understanding the processes underpinning thermal transport in such mesoscopic structures has significant potential to further our understanding of quantum thermodynamics \cite{annurev-physchem-040513-103724,Alicki_1979,PhysRevE.76.031105,polini2022materials,Myers:2022lvm} and for applications in quantum information devices, for example in the circuit's heat management \cite{Tan2017,Partanen2018}. Superconducting circuits present a practical, controllable platform in which to realise such quantum thermal devices \cite{RevModPhys.78.217, RevModPhys.93.041001}. Josephson-junction elements form quantum bits (qubit) or multi-level systems that can be strongly and controllably tuned to interact with microwave photons stored in a superconducting resonator \cite{Blais2004, Nakamura1999}. The inclusion of resistive normal-metal elements in the resonator, whose electronic temperature can be controlled and monitored, provides sources of thermal photons and, acts as a sensor of transferred power. The field's focus ranges from photonic heat transport, operation of quantum heat engines and refrigerators and calorimetry of single photons.

Rapid development in the field of cQTD have seen the realisation of practical heat transport devices, advancing our understanding of quantum thermodynamics. Experiments have measured photonic heat flow between two resistors through a superconducting quantum interference device (SQUID) in various configurations, indicating quantum limited photonic thermal conductance  \cite{Timofeev2009,Meschke2006,Giazotto2012, Fornieri2017}. This quantum-limited heat conduction is also observed across two resistors separated by 1 meter transmission line long distance \cite{Partanen2016}. Further efforts saw studies of the heat flux mediated by a qubit embedded between two microwave cavities. By utilising symmetric and asymmetric resonators, this led to the realisation of the `quantum heat valve' (QHV)  \cite{Ronzani2018a} and `quantum heat rectifier' (QHR) \cite{Senior2020} respectively. More recently, by coupling a third microwave cavity to a flux qubit, heat transport in a three-terminal device has been realised \cite{Gubaydullin2022}.

As the field of cQTD matures and is advancing, it becomes increasingly important for experimentalists to have the practical tools they need to accurately design the next generation of quantum heat devices. Until now, models of quantum heat transport have focused on so-called `lumped-element' approximations \cite{Gubaydullin2022, Thomas2019}, treating structures as ideal rather than considering a specific full geometry. Furthermore, in the limit of strong coupling to the dissipative elements, the effects of coherence are suppressed and circuits can be modelled using linearised circuit elements, with remarkable success \cite{Thomas2019}.

%As the field of cQTD matures, and looks towards greater reliability and reproducibility, it becomes increasingly important for experimentalists to have the practical tools they need to accurately design the next generation of quantum heat devices. Until now, models of quantum heat transport have focused on so-called `lumped-element' approximations \cite{Gubaydullin2022, Thomas2019}, treating structures as ideal rather than considering a specific full geometry. Furthermore, in the limit of strong coupling to the dissipative elements, the effects of coherence are suppressed and circuits can be modelled using linearised circuit elements, with remarkable success \cite{Thomas2019}.

In this work, we present a guide towards simulating heat transport in cQTD platforms employing the finite-element method (FEM) within the software package Sonnet \cite{Sonneti}. Sonnet is a software package which can solve electromagnetic propagation in planar structures using a finite element method, and is widely utilised to design superconducting circuits. For example, it has been used to efficiently determine the quality factor and resonance frequency of a superconducting micro-resonator \cite{Wisbey2014}, simulating radiation loss in a superconducting circuit sensor \cite{Endo_2020}, and for designing an on-chip superconducting filter \cite{Hao2014,PhysRevApplied.17.064022}. Inspired by the use of FEM methods in designing quantum processing units in the field of quantum information processing, we describe the first applications of such techniques to the design of quantum thermal hardware. More specifically, we simulate a QHV device using a FEM method, and precisely predict the expected heat currents. We go on to compare our results to the distributed microwave circuit theory. In the future, much more complicated systems are expected to exist to realise such quantum heat engines. 

\begin{figure}
\includegraphics{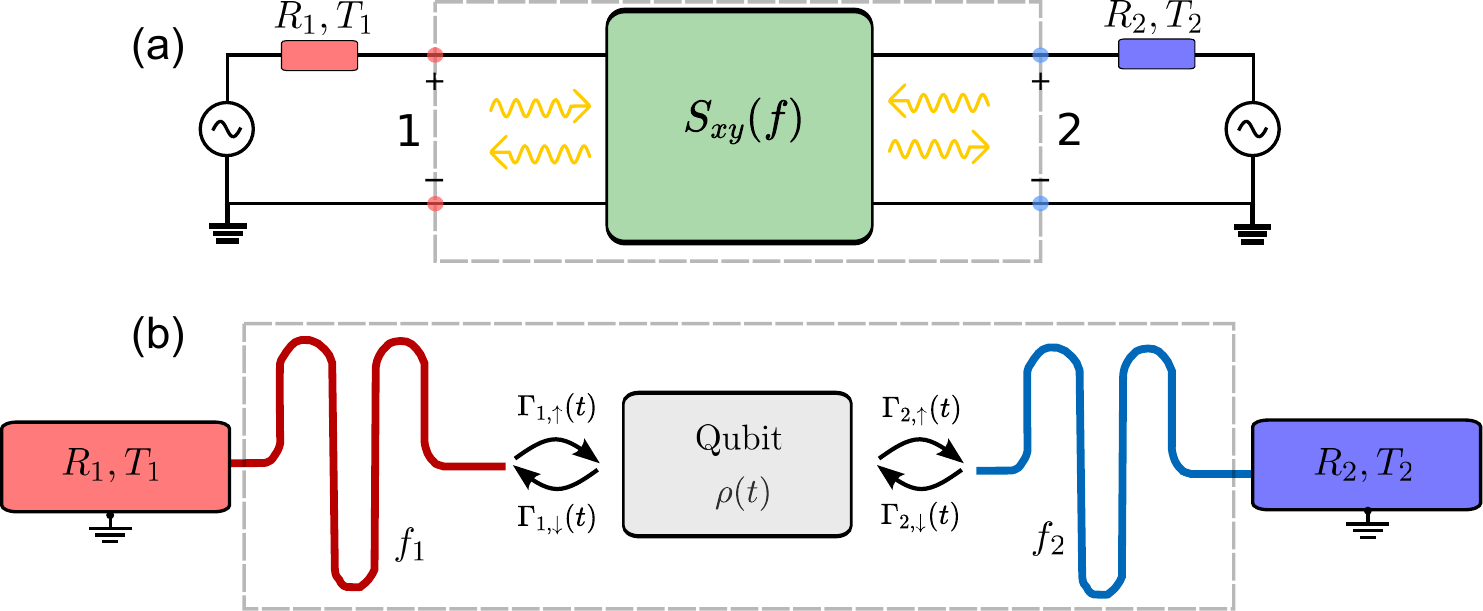}
\caption{(a) Generalised two-port thermal device consisting of a linear circuit, characterized by scattering parameter $S_{xy}(f)$, shorted at the two ends by resistors $R_1$ and $R_2$ with temperatures $T_1, T_2$. Heat is then exchanged between the two resistors via microwave photons generated by the Johnson-Nyquist noise in the resistors. (b) A schematic diagram of a typical quantum heat transport experiment, consisting of a superconducting qubit coupled by transition rates to two superconducting resonators, subsequently shorted to ground by two resistors. In the linear regime, the circuit can be represented by scattering parameters $S_{xy}$ which can be obtained from simulations. \label{fig:1}}
\end{figure}   

\section{Heat Transport through a Linear Circuit}

%Here we consider a generalised two-port thermal device, and aim to characterise such a structure within the framework of an FEM simulation.

Here we consider a generalised two-port thermal device to compute the heat transport across the device. The internal structure of the device can in-principle contain any combination of qubits, resonators, on-chip filters, capacitors, inductors, etc., which we call the `black-box', situated between an input and output microwave-port that are terminated by resistors $R_{1}$ and $R_{2}$. Both of them generate a source voltage spectrum, $\mathcal{S}_{V_{S_n}}(f)$, for $n=1,2$. At finite temperature due to thermal agitations \cite{PhysRev.32.97,PhysRev.32.110}, the metal resistor produces a voltage spectral density from the fluctuations given by \cite{RevModPhys.93.041001}
\begin{equation}
    \mathcal{S}_{V_{S_n}}(f) = \frac{2R_n hf}{1-e^{-hf/k_BT_n}},
\end{equation}
where $R_n$ and $T_n$ are the resistance and the temperature of the resistor $n$ respectively. The voltage noise $\mathcal{S}_{V_{L_2}}(f)$ accross the resistor $R_2$ is related to the input noise $\mathcal{S}_{V_{S_1}}(f)$ by the formula
\begin{equation}
    \mathcal{S}_{V_{L_2}}(f) = |H(f)|^2\mathcal{S}_{V_{S_1}}(f),
\end{equation}  
where we define the voltage transfer function, $H(f)$, as the ratio of the load voltage $V_L$ at port 2 and source voltage $V_S$ at port 1. Furthermore, the voltage transfer function can be recast in-terms of the more familiar scattering parameter ($S_{21}$), using Eq.~\ref{eq:H-to-S}, (detailed derivation is discussed in \ref{app:ABCD})
\begin{equation}
   \mathcal{S}_{P_2}(f) = \frac{1}{4R_1}|S_{21}(f)|^2 \mathcal{S}_{V_{S_1}}(f),
\end{equation}
where $\mathcal{S}_{P_2}(f) = \mathcal{S}_{V_{L_2}}(f)/R_2$ is the power-spectral density. The transmission $S_{21} = \sqrt{R_1}{V_2^{-}}/\sqrt{R_2}{V_1^{+}}$ is normalised voltage wave ratio, where ${V_1^{+}}$ and $V_2^{-}$ are the incident wave from port 1 and the total wave toward port 2, respectively. The thermal voltage spectrum is an even function, the incident power on resistor 2 from resistor 1 is then given by
\begin{equation}
\begin{split}
    P_{2} = \int_{-\infty}^{\infty}df\mathcal{S}_{P_2}(f) = \frac{1}{2R_1}\int_{0}^{\infty}df |S_{21}(f) |^2\mathcal{S}_{V_{S_1}}(f), \\
    = \int_{0}^{\infty}df h f \lvert S_{21}(f) \rvert^2 (n_1(f)+1/2),
\end{split}
\end{equation}
where $n_n(f) = 1/(e^{hf/k_BT_n}-1)$ is the Bose-function describing the thermal photon population. Using a symmetric argument for the incident power from resistor 2 back on resistor 1, along with reciprocity $S_{12} = S_{21}$, we can now write the total heat flow as
\begin{equation}
\label{eq:laundauer}
    P_\mathrm{net} = P_{2}-P_{1}= \int_{0}^{\infty}df h f \lvert S_{21}(f) \rvert^2 (n_1(f)-n_2(f)),
\end{equation}
which is a Landauer type equation \cite{PhysRevB.33.551,PhysRevLett.81.232}, where $\tau(f)=\lvert S_{21}(f) \rvert^2$ is photon transmission coefficient. We now see that solving the heat flow through an arbitrary black-box can be reduced to simply solving its scattering parameters. 

In a superconducting circuit, in the case of a QHV of Ref. \cite{Ronzani2018a}, the black-box consists of two symmetric transmission lines (TLs) capacitively coupled to a transmon qubit. Here, the terminating resistors at both ends of TLs define the boundary condition for the voltage node. Thus, the resistor-terminated TLs act as a $\lambda/4$ resonator with its open-circuit end hosting the voltage antinode to couple to the qubit. The source of microwave radiation for the QHV circuit is this normal-metal resistor shorting each $\lambda/4$ resonator to the ground-plane. The transmon qubit consists of a metal island shunted by a Josephson junction, with island total capacitance $C_{\Sigma}$ and charging energy $E_C=e^2/2C_{\Sigma}$. Here, the non-linear SQUID is replaced by an inductor $L_J$ with impedance
\begin{equation}
\label{eq:JosephsonZ}
    Z_J(\delta,\omega) = j\omega L_J(\delta)= j\omega\frac{\Phi_0}{2\pi I_{C\Sigma}\lvert\cos(\delta)\rvert\sqrt{1+d^2\tan^2(\delta)}},
\end{equation} where $\delta$, $\Phi_0$ and $I_{C\Sigma}$ are the effective phase across the SQUID, magnetic-flux quantum and total critical current of the SQUID junctions respectively. The parameter $d$ is the critical current asymmetry \cite{Koch2007}
\begin{equation}
    d = \frac{I_{C1} - I_{C2}}{I_{C1} + I_{C2}},
\end{equation}
where $I_{C1}$ and $I_{C2}$ are the critical currents of the two SQUID junctions. The inductor stores the Josephson energy $E_J(\delta)=(\Phi_0/2\pi)^2(1/L_J(\delta))$. In this linearized picture, the transmon qubit is represented as an ideal harmonic oscillator with frequency
\begin{equation}
    f_Q(\delta) = \frac{\sqrt{8E_J(\delta)E_C}}{h},
\end{equation} 
thus ignoring the in-built weak anharmonicity of the qubit. 

When the island is shunted by two parallel Josephson junctions, the phase $\delta$ is magnetic-flux dependent $\delta=\pi\Phi/\Phi_0$. Figure~\ref{fig:1}(b) shows a schematic representation of the QHV circuit, with the corresponding frequencies and rates shown. To simulate in the linear regime, we transform this to a black-box terminated by port-impedances, as shown in Fig.~\ref{fig:1}(a).

%In a Sonnet simulation, we transform these elements as the port-impedence which terminate the black-box.

\section{Determining the scattering parameter $S_{21}(f)$}

The scattering parameters of a linear circuit can be calculated by various methods. In the lumped element approximation, at low temperatures, when the thermal photon wavelength is much longer than the typical dimension of the circuit, the transmission coefficient $\tau$ between the two resistors can be derived by standard circuit approach \cite{Schmidt2004,Pascal2011}: $\tau(f)=R_1 R_2 / \lvert Z_t(f) \rvert^2$, where $Z_t(f)$ is the total series impedance of the circuit. In a typical resonator-qubit system, depending on the type and resonance frequency of the resonator, for example for $\lambda/4$ resonator with $f_r\sim8~\mathrm{GHz}$, the photon wavelength $\lambda\sim15~\mathrm{mm}$ is already comparable with the typical size of the resonator-qubit-resonator structure. This can be modelled, as in Refs. \cite{Thomas2019,Gubaydullin2022}, taking into account the distributed elements of the resonators, while still treating capacitors as a lumped element. 

Here we propose a method to solve the transmission coefficient with FEM by using Sonnet to take into account full circuit reactive elements and their possible parasitics. 
In Sonnet, for the FEM simulations, the resistive elements correspond to the port-normalising impedances which terminate the black-box. In the software, the port impedance can have an arbitrary combination of resistive and reactive elements, that can be varied to solve the transmission of the circuit (see in \ref{app:Port} for more discussion about ports in Sonnet). Here we vary only resistive elements and set the reactances to be zero. By doing this we can get the transmission of the full circuit as varying the terminating resistances. %Figure \ref{fig:1}(b shows a schematic representation of the QHV circuit, with the corresponding frequencies and rates shown.% To simulate, we transform this to a black-box terminated by port-impedences.%, as shown in \ref{fig:1}(below). 

As a benchmark, we also solve the transmission of the circuit using the individual distributed circuit elements, by constructing the ABCD matrix of the black-box and converting to its scattering parameters. The ABCD matrix of the entire circuit is then found by computing the product of the corresponding ABCD matrices of each of the constituting circuit elements \cite{Pozar:882338}

\begin{equation}
\begin{pmatrix}
A & B
\\C & D
\end{pmatrix}
=
\begin{pmatrix}
A_1 & B_1
\\C_1 & D_1
\end{pmatrix}
\begin{pmatrix}
A_2 & B_2
\\C_2 & D_2
\end{pmatrix}
\begin{pmatrix}
A_3 & B_3
\\C_3 & D_3
\end{pmatrix}...
\begin{pmatrix}
A_n & B_n
\\C_n & D_n
\end{pmatrix}.
\end{equation} This matrix can then be transformed back to the scattering parameters using the relationship (in-detail derivation discussed in \ref{app:ABCD})
\begin{equation} 
S_{21}(f)=\frac{2\sqrt{R_1/R_2}}{A+B/R_2+CR_1+(R_1/R_2)D}.
\label{eq:S21_conversion}
\end{equation}Photon transmission probability, $\lvert S_{21}(f) \rvert^2$, calculated from Eq.~\ref{eq:S21_conversion} corresponds to that of Ref. \cite{Schmidt2004,Pascal2011} when the black-box can be represented by a total series impedance, and corresponds to that of Ref. \cite{Thomas2019,Gubaydullin2022} when the black-box can be represented by a total admittance of the parallel elements (see the discussions in \ref{app:ABCD}). For example, when the ports are directly connected, without any series or parallel impedances, with port termination $R_1$ and $R_2$, the matrix elements are $A=1,B=0,C=0,D=1$. Thus the photon transmission probability $\tau(f)=\lvert S_{21}(f) \rvert^2={4R_1R_2}/{(R_1+R_2)^2}$.

%In any arbitrary linear circuit, the task now is how to solve and calculate the transmission coefficient $\lvert S_{21}(f) \rvert^2$ of a circuit under investigation.

%In the limit that the size circuit elements tends to zero we approach the case of the FEM simulation, where each small section of the circuit is modelled by an ABCD matrix and the specific geometry plays a role. Such a FEM scattering problem can be solved using a commercial package, such as Sonnet. Here we compare and contrast these methods for calculating the photonic heat transport.

\section{Heat Transport through a Superconducting Quarter-Wave Resonator}
\begin{figure}
\includegraphics{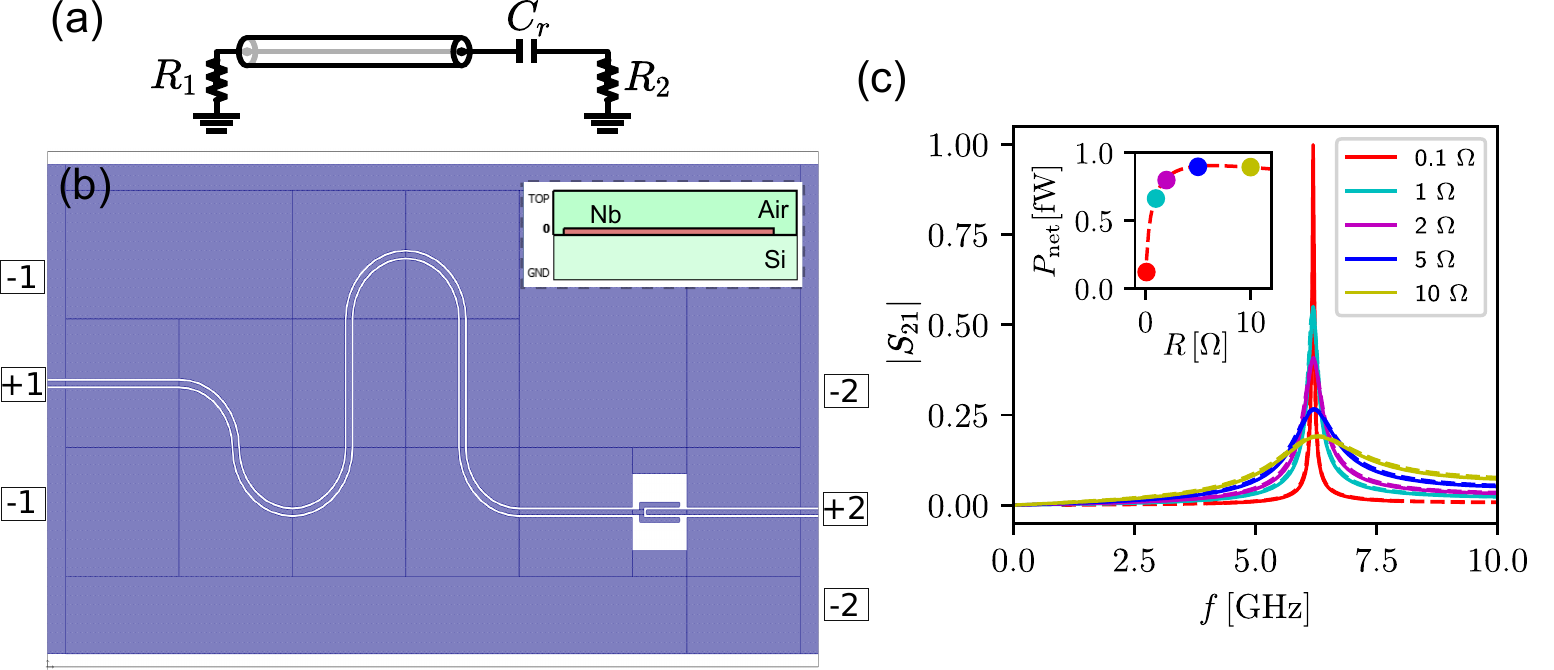}
\caption{(a) Circuit schematic of a resistor-terminated $\lambda/4$ resonator coupled to another resistor. (b) Sonnet configuration for simulating a resistor-terminated quarter-wave resonator. The green sections show the planar metal layer of superconducting Nb, and the white is the silicon dielectric. The two-ports are shown as the numbers (1 and 2) connecting the metal structures to the simulation boundary box. The sign positive (+) connects the metal to the ground with sign negative (-). (c) Comparison of simulated $|S_{21}|$ between Sonnet simulation (solid line) and ABCD-matrix analysis (dashed lines) for a range of terminating $R$ at the port 1, showing excellent agreement. The inset shows the net power to the port-2 when increasing resistance $R$ at port-1, coloured dots are calculated by Sonnet simulation method and red-dashed line from ABCD model. \label{fig:2}}
\end{figure}

To demonstrate this approach, we first consider the simple-case of heat transport through a superconducting $\lambda / 4$ resonator. The circuit consists of a resistor, $R$, at port-1 terminating a $6~\mathrm{GHz}$ $\lambda / 4$ resonator. The open end of the resonator capacitively couples to a short TL terminated by a matched $50~\,\mathrm{\Omega}$ resistor at port-2. In this way, we find the scattering parameters, $S_{21}(f)$, as a function of the terminating resistor at port-1, and eventually the total power transfer to port-2. Due to the simplicity of the circuit, ABCD methods and Sonnet can be compared directly as methods for determining the heat flow. Figure~\ref{fig:2}(a) and (b) shows the schematic and Sonnet configuration of the corresponding circuit. The presented superconducting structure is approximated to be a zero-thickness metal with perfect conductance. The dielectric stack-up consists of a vacuum layer (dielectric constant $\epsilon = 1$) above the metal layer, and a $670~\mathrm{\mu m}$ silicon ($\epsilon = 11.5$) layer below the metal with zero dielectric loss. The Sonnet simulation of the scattering parameters is then performed for a range of terminating resistors, and the results are shown as the solid-lines in Fig.~\ref{fig:2}(c). %More detail of the simulation parameters are given in \ref{App:parameters}.

The scattering parameters can be correspondingly calculated using the product of the ABCD matrices for the individual elements. The product is given by the three elements of the circuit
\begin{equation}
\begin{pmatrix}
A & B
\\C & D
\end{pmatrix}
=
\underbrace{\begin{pmatrix}
\cos{\beta l_1} & jZ_0\sin{\beta l_1}
\\j\frac{1}{Z_0}\sin{\beta l_1} & \cos{\beta l_1}
\end{pmatrix}}_\textrm{Transmission line}
\underbrace{\begin{pmatrix}
1 & \frac{1}{jC_r \omega}
\\0 & 1
\end{pmatrix}}_\textrm{Coupling Capacitor}
\underbrace{\begin{pmatrix}
\cos{\beta l_2} & jZ_0\sin{\beta l_2}
\\j\frac{1}{Z_0}\sin{\beta l_2} & \cos{\beta l_2}
\end{pmatrix}}_\textrm{Transmission line},
\end{equation}
where $Z_0$ is the characteristic impedance of the transmission line, $l_1$ and $l_2$ are the lengths of the two transmission line sections, $l_1 > l_2$, and $\omega$ is the input frequency. In previous experimental results \cite{Chang2019}, internal loss ($1/Q_i$) to the substrate has been observed to be very small compared to the loss to the resistor ($1/Q_R$), i.e. photons mostly decay to the resistor. We therefore set the attenuation constant to zero, and $\beta =\omega l \sqrt{C_l L_l}$. The resultant product is converted to $S_{21}$ using Eq.~\ref{eq:S21_conversion}, and shown as the dashed-lines in Fig.~\ref{fig:2}(b) demonstrating excellent agreement between the FEM and ABCD methods for all values of resistances. The total integrated power transfer, calculated using Eq.~\ref{eq:laundauer}, as a function of resistance $R$ at port-1 is shown in the Fig.~\ref{fig:2}(b) inset. 

In the case of this simple circuit the role of parasitic couplings and modes are minimised, and the scattering parameters are well approximated by the ABCD matrices of the individual components. As circuits become increasingly complex, parasitic capacitances and inductances can no-longer be neglected and the ABCD approach is expected to diverge from the true circuit response. A major advantage however, is that the Sonnet simulation is performed without recourse to any knowledge about the circuit components, only inputting the design file and desired resistance. Conversely, the ABCD method requires the additional steps of simulating the coupling capacitance, and characteristic impedance using an external program.

\section{Quantum Heat Valve: A qubit coupled to two superconducting resonators}
\begin{figure}
\includegraphics{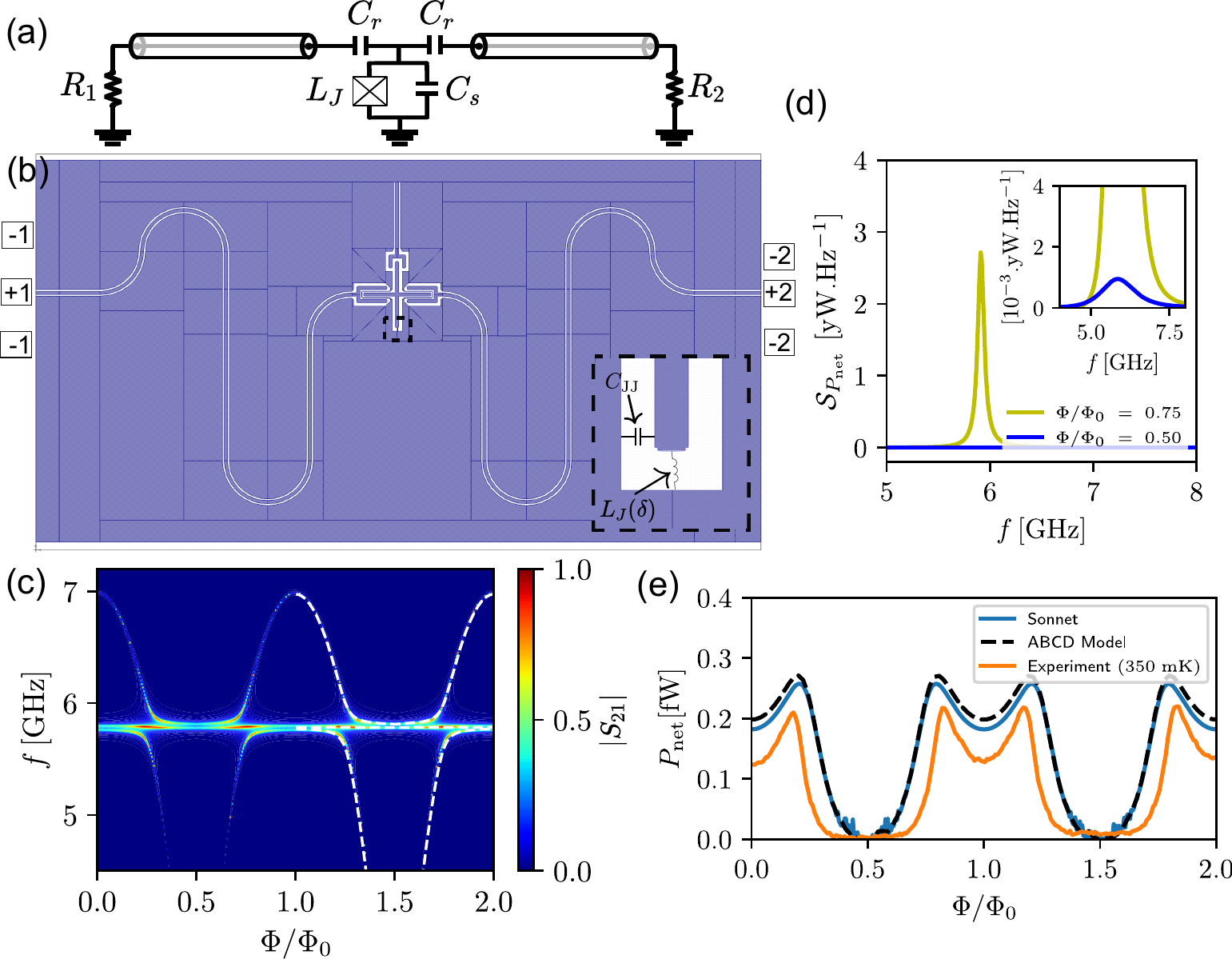}
\caption{(a) Circuit schematic of a transmon qubit coupled to two equal $\lambda/4$ resonators. (b) Sonnet configuration for simulating a QHV device, with the transmon qubit and two superconducting transmission lines shorted by a resistor, forming $\lambda/4$ resonators. The zoomed image shows the transmon qubit, with associated inductor, used to simulate a linearised Josephson junction. (c) Spectroscopy of sonnet-simulated $|S_{21}|$ at different flux of the device with low terminating resistance $R_1 = R_2 = 0.1~\mathrm{\Omega}$. White dashed-lines are fits according to the Hamiltonian given in Eq.~\ref{eq:HamiltonianQHV}. (d) One-sided total power spectrum at resistor 2, integrand of Eq.~\ref{eq:laundauer}, at open-valve (yellow) and closed-valve (blue). (e) Simulated heat flux calculated using Eq.~\ref{eq:laundauer}, for the Sonnet simulation method (solid blue line), and for comparison using the ABCD (black dashed-line) for a resistor 1 temperature, $T_1 = 350~\mathrm{mK}$) and resistor 2 temperature, $T_2 = 120~\mathrm{mK}$. The solid orange line is the experimental data taken from Ref.~\cite{Ronzani2018a} at the same nominal temperatures. For comparison, the unmodulated background has been removed from the experimental data. \label{fig:3}}
\end{figure}

Having demonstrated the validity of the linear FEM simulations to simulate heat flow, we move to the more complex case of the QHV, inspired by the experimental work \cite{Ronzani2018a}. The QHV consists of a superconducting transmon qubit, coupled to two superconducting  $\lambda/4$ resonators of equal frequency. The transmon qubit frequency is tunable using a global flux bias to modulate the Josephson inductance of a superconducting-quantum-interference-device (SQUID). We approximate the transmon qubit, considering only the linear response, by replacing the SQUID loop with an ideal lumped inductor within the Sonnet interface. Figure~\ref{fig:3}(a) and (b) show the circuit schematic and Sonnet setup for such simulations, with the inset showing the tunable inductor representing the transmon SQUID. Ports are placed at each of the short-ends of the $\lambda$/4 resonators, and the port impedance set to the desired resistor value. The metallic layer is assumed to be lossless and have zero intrinsic inductance, and the ground planes are connected to the box-wall such that the impedance to ground is zero at the boundary. Additionally, a small $C_\mathrm{JJ} = 10~\mathrm{fF}$ capacitor is added between the transmon island and the ground-plane, to account for the $0.2~\mathrm{\mu m^2}$ area junction capacitance.

The Josephson inductance is calculated by Eq.~\ref{eq:JosephsonZ}, for each value of the simulated phase, $\delta=\pi\Phi/\Phi_0$. The S-parameters are simulated using Sonnet. The results of a typical simulation as a function of flux are shown in the colour axis of Fig.~\ref{fig:3}(c), with $R_1 = R_2 = 0.1~\mathrm{\Omega}$ for visual clarity. The interaction of the qubit with the two resonators is shown clearly by the two avoiding crossings occurring each period. By fitting the eigenenergies using the SCQubits package \cite{Groszkowski2021scqubitspython}, shown by the white dashed lines (Eq.~\ref{eq:HamiltonianQHV}), we can further extract the qubit-resonator coupling $100~\mathrm{MHz}$, and charging energy $E_c / h = 147~\mathrm{MHz}$ in excellent agreement with the experimental value $150~\mathrm{MHz}$.

To compute the power transferred, we first convert the $S_{21}$ to the one-sided net power transfer spectrum at the second resistor by $\mathcal{S}_{P_{net}} = h f \lvert S_{21}(f) \rvert^2 (n_1(f)-n_2(f))$, shown in Fig.~\ref{fig:3}(d) for two values of the flux. Note, that we set the port resistance to $R_1 = R_2 = 12~\mathrm{\Omega}$, corresponding to a quality factor $Q_1 = Q_2 = 3.1$, matching the fitted experimental values. The yellow solid line indicates the power spectra when the valve is in the open position, and the blue line when the valve is in the closed position. The inset shows a zoom of the data when the valve is in the closed position. The Lorentzian shape is therefore created by the spectral filtering of the resonators around $5.6~\mathrm{GHz}$. The total power transferred is then naturally obtained by integrating the power spectral density over all frequencies. The power as a function of flux for three temperature bias values is shown as the solid lines Fig.\ref{fig:3}(e). For all curves, the temperature of the drain-side is fixed at $T_2 = 120~\mathrm{mK}$. 

For comparison, we again compute the scattering parameters using a linearised ABCD product of each corresponding element. The product is given by
\begin{equation}
\begin{split}
\begin{pmatrix}
A & B
\\C & D
\end{pmatrix}
=
&
\underbrace{\begin{pmatrix}
\cos{\beta l} & jZ_0\sin{\beta l}
\\j\frac{1}{Z_0}\sin{\beta l} & \cos{\beta l}
\end{pmatrix}}_\text{Transmission Line}
\underbrace{\begin{pmatrix}
1 & \frac{1}{jC_r \omega}
\\0 & 1
\end{pmatrix}}_\text{Coupling Capacitor}
\underbrace{\begin{pmatrix}
1 & 0\\
\frac{Z_C+Z_J}{Z_CZ_J} & 1
\end{pmatrix}}_\text{Qubit}
\\&
\underbrace{\begin{pmatrix}
1 & \frac{1}{jC_r \omega}
\\0 & 1
\end{pmatrix}}_\text{Coupling Capacitor}
\underbrace{\begin{pmatrix}
\cos{\beta l} & jZ_0\sin{\beta l}
\\j\frac{1}{Z_0}\sin{\beta l} & \cos{\beta l}
\end{pmatrix}}_\text{Transmission Line},
\end{split}
\end{equation}
where $Z_C$ and $Z_J$ are the lumped impedances representing the qubit shunting capacitance ($C_s$), and Josephson inductance respectively (parallel LC circuit). The calculated circuit model is shown as the dashed line in Fig.~\ref{fig:3}(d) for comparison. To compare, the coupling capacitances and qubit charging energy are simulated using COMSOL. The Josephson energy, $E_J = 37~\mathrm{GHz}$ ($I_{C\Sigma} = 72~\mathrm{nA}$) and critical current asymmetry $d = 0.08$ are taken to be the same in both models. Total power transfer is again calculated by integrating the simulated $S_{21}$ over the full frequency range using Eq.~\ref{eq:laundauer}. The two models demonstrate in general excellent quantitative agreement. However at integer values of the flux quanta, where the power transfer is maximised, there is some discrepancy between the models. In general, the circuit method using COMSOL capacitance values overestimates the power transfer compared to the more precise Sonnet method. We attribute this discrepancy to a more accurate estimation of the qubit-resonator coupling by Sonnet when compared to COMSOL. This is because the coupling capacitor in Sonnet is treated as a distributed element whilst in the ABCD model it is assumed to be a lumped element. 

Comparing to the experimental data from Ref. \cite{Ronzani2018a}, shown by the solid orange line, we find excellent qualitative agreement, suggesting that the linearised model simulates the dynamics well. The measurements observed an overall lower peak power modulation of $\Delta P_\mathrm{net} = 0.21~\mathrm{fW}$, versus the simulated $\Delta P_\mathrm{net} = 0.29~\mathrm{fW}$ for the same nominal experimental parameters. The observed discrepancy comes partly from the non-linearity caused by the weak anharmonicity of the transmon qubit, and as such the populations of the quantized energy levels play a non-negligible role in filtering the power-transfer in such experiments. Alternatively, elements of the fabrication, or measurement environment, e.g. sample holder, measurement wiring and wirebonding, can play a role in determining the overall magnitude of the heat flow, something we will further explore. Overall, the close agreement obtained between the experiment and the simulations is remarkable considering the simplified model, and lack of free parameters when constructing the simulation.

Sonnet simulations allow quantitative estimations of the background heat flow due to photons in superconducting circuits. By looking at the off-resonant heat flow ($\Phi / \Phi_0 = 0.5$) we can observe that net power flow is almost zero when compared with the resonant heat flow. In-fact, we calculate the modulation ratio ($P(\Phi)_\mathrm{max} - P(\Phi)_\mathrm{min} /P(\Phi)_\mathrm{max}$) from the Sonnet simulations to be $0.95 \pm 0.02$, in stark contrast to that seen in recent experimental results. From this we would conclude that the majority of the observed background heat flow in experiments is due to phonons, which are not considered by Sonnet. However, the picture can become more complex when we consider the possible variation or grounding potential of the measurement environment. Here so far we simulate the circuit in the ideal situation where the ground plane of the circuit is connected to the box-wall.

%\section{Suppressing the Photonic Background Contribution} %\label{sec:background}

\begin{figure}
\includegraphics{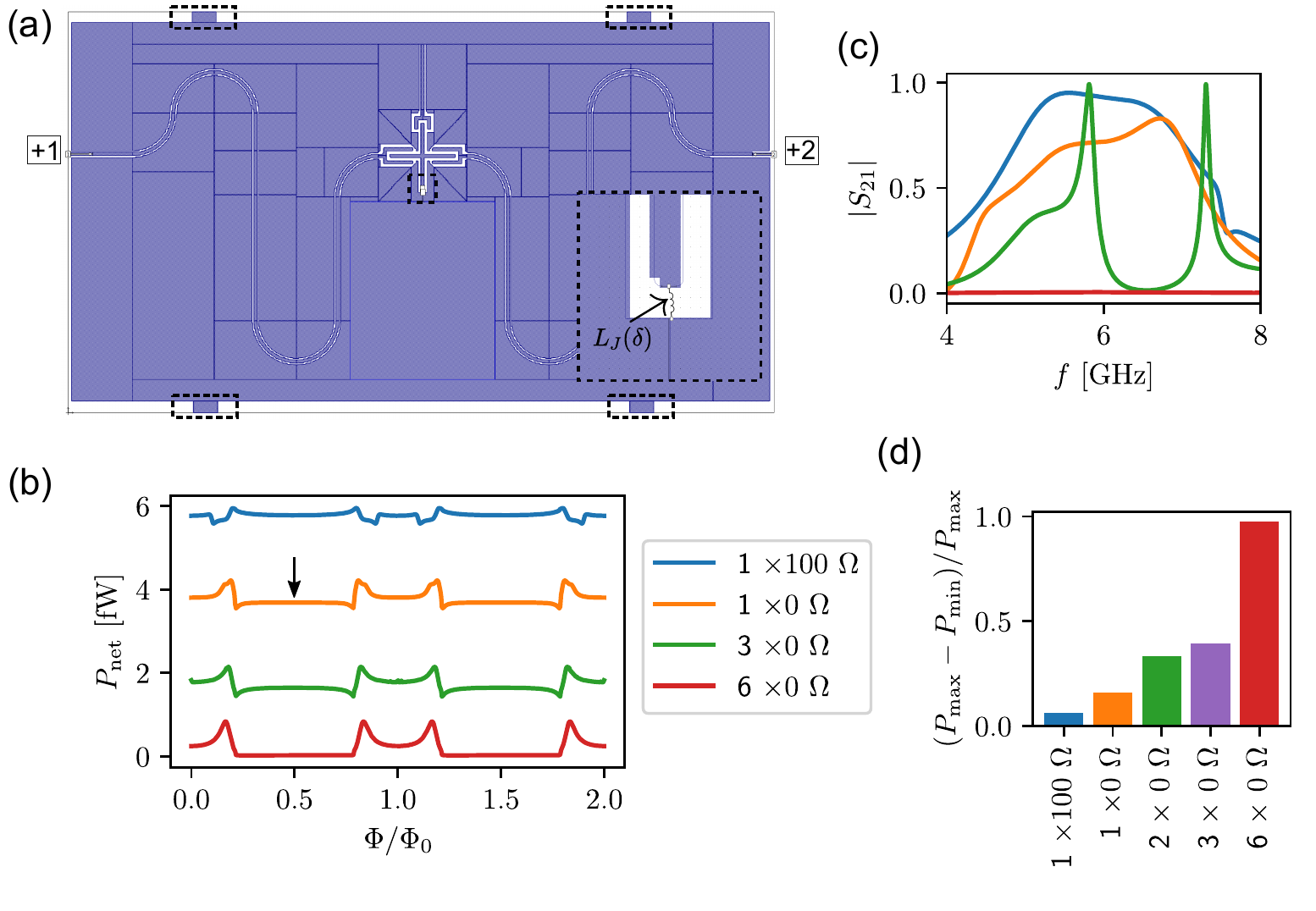}
\caption{(a) Sonnet setup for investigating the effect of wire-bonding on the photonic background contribution. The layout is a simplified QHV device consisting of a transmon qubit between two superconducting resonators. Bonds are simulated by connecting the ground plane to the grounded box-wall (four shown here), either by lossless metal, or through lumped resistors. (b) Simulated heat-flow as a function of flux for four different bonding configurations. It is shown clearly that an increased impedance between the ground-plane and circuit ground contributes significantly to the background heat and the shape of the modulated heat flow. The black arrow indicates the flux point chosen for plot (c). (c) Off-resonant $S_{21}$ transmission for four different circuit configurations. Fewer wire-bonds allows the propagation of parasitic modes at low-frequency, evidenced by the strong transmission around $6~\mathrm{GHz}$ for the blue and orange curves. (d) Modulation ratio, as defined by $(P(\Phi)_\mathrm{max} - P(\Phi)_\mathrm{min}) /P(\Phi)_\mathrm{max}$ for five different bonding configurations. \label{fig:gnd}}
\end{figure}

To further explore how the measurement environment can affect the unmodulated background in such  circuits we consider a similar QHV device in a variety of measurement configurations. We realise this by altering the connections from the circuit ground-plane to the so-called `box-wall', which sets the simulation ground potential. This allows us to simulate the real effect of various measurement configurations. Figure~\ref{fig:gnd}(a) shows such a simulation configuration with the ground-plane short to the box-wall using four lossless connections, emulating for example four superconducting wire bonds directly to the sample-holder ground. Note that the qubit coupler design is simplified with-respect-to Fig.~\ref{fig:3} to allow for faster simulation.

The Sonnet simulations here point to a clear effect of an imperfect measurement environment on the photonic heat-flow. Figure~\ref{fig:gnd}(b) shows the integrated heat-flow between the two-resistors as a function of flux, for four different measurement environments. The red, green and orange curves show the effect of an increasing number of zero-resistance wire-bonds to the chip. The blue curve represents a grounding connection through a high-impedence DC-line. Three effects are made clear: firstly, an increased impedence to ground contributes to a higher off-resonant heat flow, evidenced by the increase in the background heat flow. Secondly, the absolute magnitude of the modulation is also affected, with $\Delta P_\mathrm{net}$ reducing 20\% as the number of bonds is reduced from six to one. Lastly, the apparent shape of the modulation is also influenced, with the peak caused by the qubit interaction reducing due to competition with the background modes. In the extreme case of the blue curve, the total heat-flow is highest when the qubit is off-resonance, a $\pi$-phase shift of the QHV characteristics.

The source of this behaviour is clear when we look at the off-resonance ($\Phi = 0.5$) $|S_{21}|$ transmission for the different cases, as shown in Fig.~\ref{fig:gnd}(c). With fewer connections, the ground-plane allows for the propagation of significant background modes, seen increasing in amplitude from the green, orange and blue curves. Note that the exact background modes and their amplitude depend significantly on the physical position of the bonds on the chip. The interaction between the tunable QHV modes and the parasitic modes results in the phase shift of heat-valve behaviour. Moreover, the increased background results in a reduced modulation ratio, as seen in Fig.~\ref{fig:gnd}(d).

%Interestingly, relatively few bond-wires are needed to suppress this behaviour as shown by the red curve representing just six bonded connections.

Simply changing the measurement environment can lead to an order-of-magnitude reduction in the modulation ratio, although the absolute modulation is left unaffected. This cements the importance of maintaining a precise environment in the measurements in order to study the quantum thermal device performance. Such effects may shed further light on some recent experimental results which report modulation which could not be easily explained within a circuit framework \cite{Senior2020, Gubaydullin2022}.

%Building upon this, together with an understanding of the applicability of our simulation techniques, we now apply them to the design of a future photonic heat device.

%It is important to note here that the exact heat current depend also on the spatial distribution of the ground connections as well as the number.

%This cements the importance of maintaining a constant environment in-order-to draw consistent conclusions about the device performance.

\section{Double Pole Quantum Heat Valve: Two qubits between two superconducting resonators}

\begin{figure}
\includegraphics{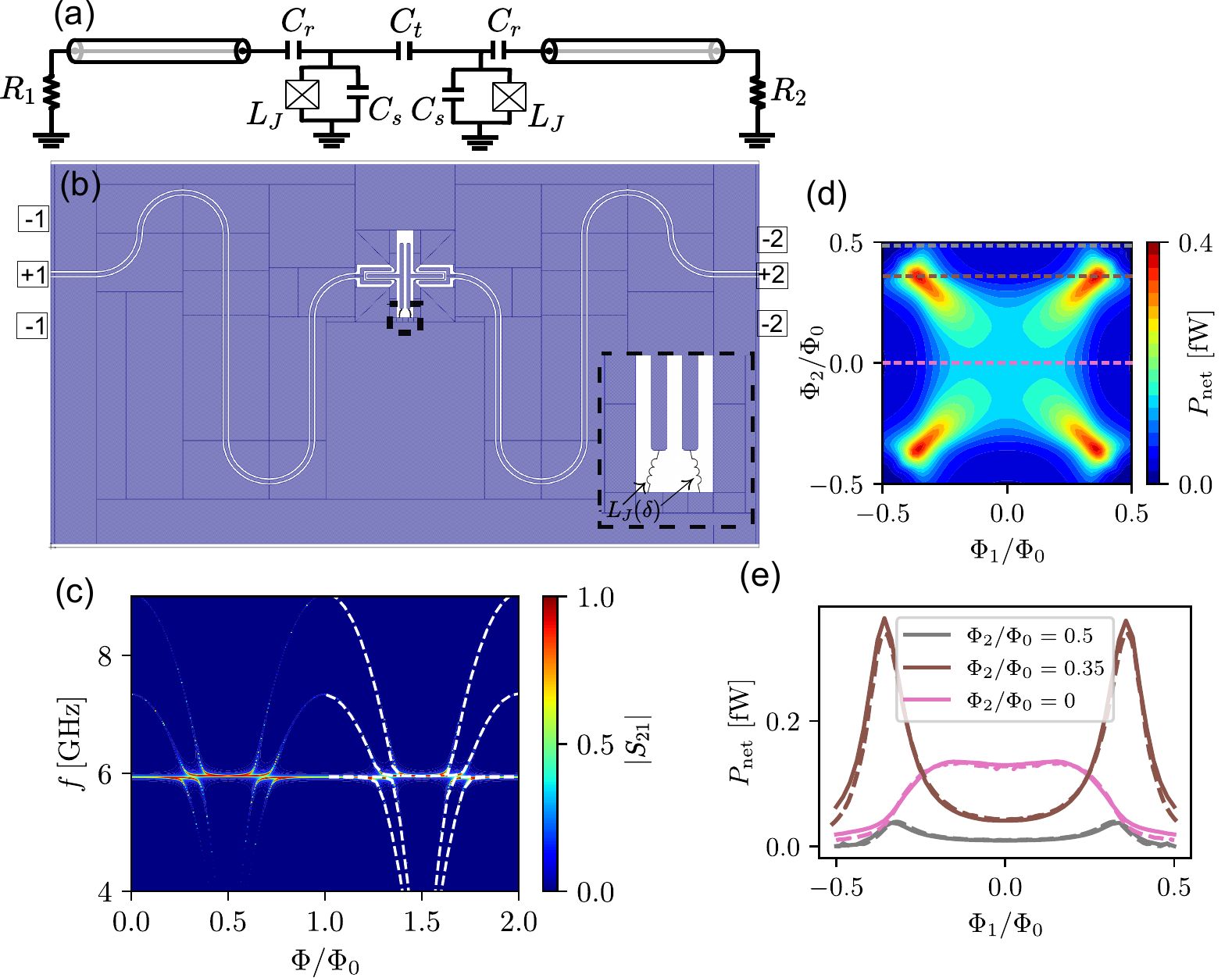}
\caption{(a) Circuit schematic of two transmons coupled to two $\lambda/4$ resonators. (b) Sonnet configuration for simulating a double pole QHV device, with the two transmon qubits and two superconducting $\lambda/4$ resonators labelled. The zoomed image shows the two transmon qubits, with associated inductors, used to simulate values of a linearised Josephson junction. (c) Spectroscopy of Sonnet-simulated $|S_{21}|$ at different flux values of the device with low terminating resistance $R_1 = R_2 = 0.1~\mathrm{\Omega}$. White dashed-lines are fits according to the Hamiltonian given in Eq.~\ref{eq:Hamiltonian2QHV}. (d) Simulated heat flux calculated using Eq.~\ref{eq:laundauer} from the Sonnet simulation, as a function of qubit 1 flux ($\Phi_1$) and qubit 2 flux ($\Phi_2$). (e) Three curves (solid lines) of the simulated heat flux obtained using slices of (d). The resistor 1 temperature, $T_1 = ~350~\mathrm{mK}$ while the resistor 2 temperature $T_2 = 120~\mathrm{mK}$. The dashed lines are power calculated from ABCD-matrix model.\label{fig:5}}
\end{figure}

With the methods well established, we can now use our toolbox to design the next generation of quantum heat devices. One example of this could be a double-pole quantum heat valve. The QHV can be further expanded upon by replacing the single qubit with two strongly-coupled transmon qubits. The device therefore consists of two quarter-wavelength resonators of equal frequency $5.6~\mathrm{GHz}$, each coupled to an transmon qubit, which are strongly coupled to each other. The schematic, and device layout in Sonnet are shown in Fig.~\ref{fig:5}(a) and (b). The charging energies and Josephson energies of the two qubits are designed to be equal. The two resonant frequencies corresponding to these qubits can be characterised by a global flux bias. Conversely, by using local flux biases the frequencies of the qubits can be tuned independently, and a two-pole photonic heat switch can be realised. Such a device serves as a building-block towards complex logic involving photonic heat currents, since it converts two inputs to a single output. 

The simulation result in Fig.~\ref{fig:5}(c) when the two qubits are tuned with equal flux, and shows a mode structure of the device is similar to the QHV. Here we set the same Josephson energy $E_J = 37~\mathrm{GHz}$ ($I_{C\Sigma} = 72~\mathrm{nA}$) for both qubits, which implies that they also have the same resonant frequency at all values of the flux bias, therefore the coupled qubits form hybridised modes. The resonator mode is identical with the single-qubit QHV, but instead of a single frequency qubit mode, the strong coupling between the qubits splits the shared resonance frequency into two. Again, using the SCQubits package we can extract the device parameters directly from the $S_{21}$ simulation. We find the qubit charging energy to be $E_{C}/h = 250~\mathrm{MHz}$, the  qubit-qubit coupling to be $g_{\alpha \beta} = 200~\mathrm{MHz}$, and the qubit-resonator coupling $g_{1\alpha} = g_{2\beta} =120~\mathrm{MHz}$. The cross-coupling terms $g_{12} = g_{1\beta} = g_{2\alpha} \sim 0$ within the fitting error.

We determine the heat current in this two-qubit device using Eq.~\ref{eq:laundauer}, as a function of the flux applied to each of the qubits, as shown by the colour axis in Fig.~\ref{fig:5}(d). Four high power peaks are seen when both qubits are tuned close to the resonator at $~\Phi_i/\Phi_0 \approx \pm 0.4$. As expected, if either qubit is detuned to a half-integer flux point, then the power remains small over the full flux range of the other qubit. In this way, the system is acting as a double-pole heat switch. The 1D slices of the 2D data corresponding to three values of the second qubit flux, indicated by the dashed lines, are shown in Fig.~\ref{fig:5}(e). We compute the ABCD product of the device as
\begin{equation}
\begin{split}
\begin{pmatrix}
A & B
\\C & D
\end{pmatrix}
=
&
\underbrace{\begin{pmatrix}
\cos{\beta l} & jZ_0\sin{\beta l}
\\j\frac{1}{Z_0}\sin{\beta l} & \cos{\beta l}
\end{pmatrix}}_\text{Transmission Line}
\underbrace{\begin{pmatrix}
1 & \frac{1}{jC_r \omega}
\\0 & 1
\end{pmatrix}}_\text{Coupling Capacitor}
\underbrace{\begin{pmatrix}
1 & 0\\
\frac{Z_C+Z_J}{Z_CZ_J} & 1
\end{pmatrix}}_\text{Qubit}
\\&
\underbrace{\begin{pmatrix}
1 & \frac{1}{jC_t \omega}
\\0 & 1
\end{pmatrix}}_\text{Coupling Capacitor}
\underbrace{\begin{pmatrix}
1 & 0\\
\frac{Z_C+Z_J}{Z_CZ_J} & 1
\end{pmatrix}}_\text{Qubit}
\underbrace{\begin{pmatrix}
1 & \frac{1}{jC_r \omega}
\\0 & 1
\end{pmatrix}}_\text{Coupling Capacitor}
\underbrace{\begin{pmatrix}
\cos{\beta l} & jZ_0\sin{\beta l}
\\j\frac{1}{Z_0}\sin{\beta l} & \cos{\beta l}
\end{pmatrix}}_\text{Transmission Line}.
\end{split}
\end{equation}
Similar to the previous comparisons, we estimate the capacitances with COMSOL and convert the ABCD matrix to the S-parameter $S_{21}$, which is then integrated according to Eq.~\ref{eq:laundauer}. The models again show excellent agreement over the full flux range. Such a device could be practically realised using current fabrication and measurement techniques. Furthermore, it could serve as a test-bed for investigating the effects of qubit coherence on heat-flow \cite{Cattaneo2021}.

\section{Conclusions}

We have demonstrated the first applications of FEM simulations to improve the design of photonic heat devices and calculate heat transport in superconducting circuits. We first established the technique and theory, showing that such simulations can calculate the scattering parameters of an arbitrary geometry, and predict the expected heat transport properties. We use our tools to predict the heat current across a simple quarter-wavelength resonator terminated by a normal-metal resistor, finding excellent agreement with established circuit models. We then predicted the heat currents at various temperatures in a QHV device, consisting of a transmon qubit coupled to two quarter-wavelength resonators, finding quantitative agreement within ~30\% of experimental data. 

We show that Sonnet can naturally predict and include any unwanted parasitic modes in the calculations. The ability to consider the specific geometry is highly useful to design further more complex quantum heat transport devices. This is clearly evidenced by the strong dependence of the photonic heat background on the simulated measurement environment, which has been investigated. We show that the electrical environment can influence not just the magnitude of the power transfer, but can even reverse the properties of the tunable heat valve. We go on to utilise our tool to design a more complex two-pole heat valve using two transmon qubits. Such a structure has not been previously realised, and presents a step towards realising logical operations using photonic heat currents.

Moreover, the technology shown here can easily be extended to an arbitrary number of heat-baths by including more ports, allowing predictions to be made about structures with four or more ports.  Our framework is currently limited by the linearity of the Sonnet FEM method. In the future, by combining non-linear solvers \cite{9387129} with FEM simulations one could, in principle, model superconducting qubits with greater accuracy than is done here. Using such solvers, one could perhaps create heat rectifiers, isolators and circulators using FEM as the core design tool. The toolbox we establish here lays the foundations for rapid prototyping of new photonic heat devices, and allows the field of cQTD to move towards increased complexity and reproducibility.

\section*{Contributions and Acknowledgements}
A.G. and C.D.S. conceived the study idea. The simulation platform and theoretical ideas were developed and performed by I.M. and C.D.S. Data analysis was performed by A.G., C.D.S. and I.M. Figures were made by A.G. and C.D.S. All authors contributed equally to the writing of the manuscript. J.P. supervised the authors at all stages of the project.

We acknowledge Dr. Yu-Cheng Chang, Dr. Dmitry Golubev and Dr. George Thomas for technical support and insightful discussions. We thank Dr. Alberto Ronzani for providing us with the raw data for Ref.~\cite{Ronzani2018a}. This work is financially supported through the Foundational Questions Institute Fund (FQXi) via Grant No. FQXi-IAF19-06, Academy of Finland grants 312057 and from the European Union’s Horizon 2020 research and innovation programme under the European Research Council (ERC) (Grant No. 742559). We acknowledge the provision of facilities OtaNano - Low-Temperature Laboratory of Aalto University to perform this research.

\section*{Data Availability Statement}
All data used in this paper are available upon request to the authors, including descriptions of the data sets, and scripts to generate the figures. Additionally, an open-source python package called `picoQuantum', used to generate the ABCD model curves, is available for installation via pip.

\appendix

\section{ABCD Matrix to $H(f)$ and $S_{21}(f)$}\label{app:ABCD}
\begin{figure*}[h!]
\centering
\includegraphics{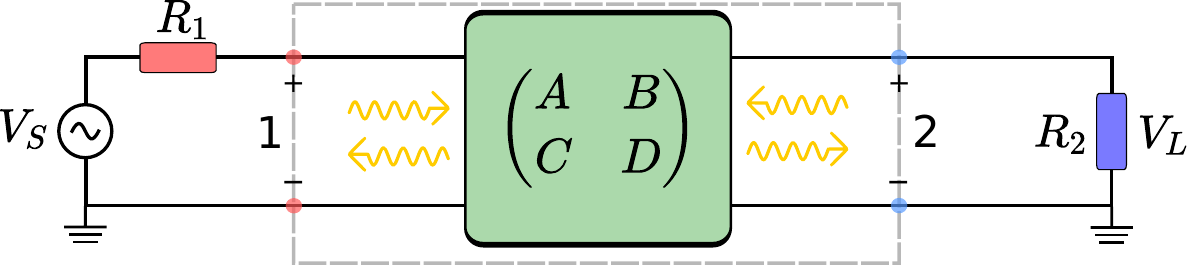}
\caption{A two-port device consisting of a linear circuit, characterised by an ABCD matrix, shorted at both ends by resistors $R_1$ and $R_2$. The voltage noise from resistor $R_1$ is modelled by a series voltage source $V_S$, and the corresponding load voltage across $R_2$ is $V_L$.  \label{fig:A1}}
\end{figure*}

The transfer function $H(f)$ can be represented in terms of the ABCD parameters by applying Kirchoff's voltage law and the definition of the ABCD matrix to the circuit shown in Fig. \ref{fig:A1}. First, by Kirchoff's voltage law:
\begin{align}
\begin{split}
 V_L &= V_2 = I_2R_2,
\\
V_S &= I_1R_1 + V_1,
\end{split}
\end{align}
where $V_{i}$ and $I_{i}$, $i \in \{1,2\}$, are the voltage and the current at node $i+$.

Second, by the definition of the ABCD matrix:
\begin{align}
\begin{split}
V_1 = AV_2 + BI_2,
\\
I_1 = CV_2 + DI_2.
\end{split}
\end{align}
Hence,
\begin{equation}
\label{eq:ABCD-to-H}
     H(f) = \frac{V_L}{V_S} = \frac{I_2R_2}{R_1I_1 + V_1}= \frac{R_2}{AR_2 + B + CR_1R_2 + DR_1}.
\end{equation}

To derive a representation for the S-parameter $S_{21}(f)$, we first calculate the input impedance

\begin{equation}
    Z_{\mathrm{in}} = \frac{V_1}{I_1} = \frac{A+B/R_2}{C+D/R_2},
\end{equation}
from which we get the reflection coefficient
\begin{equation}
    S_{11} = \frac{Z_{\mathrm{in}}-R_1}{Z_{\mathrm{in}}+R_1}
    =
    \frac{A + B/R_2 - CR_1 - D(R_1/R_2)}{A + B/R_2 + CR_1 + D(R_1/R_2)}.
\end{equation}
The voltage $V_1$ can now be written in the form:
\begin{equation}
    V_1 = V_1^{-}+V_1^{+} = V_1^{+}(1+S_{11}),
\end{equation}
where $V_1^{+}$ and $V_1^{-}$ are the incident and the reflected component respectively. Now, we can write the S-parameter

\begin{equation}
    S_{21} = \sqrt{R_1/R_2}\frac{V_2^{-}}{V_1^{+}}\biggr\rvert_{V_2^+=0}
    = \sqrt{R_1/R_2} \frac{V_2}{V_1}(1+S_{11}),
\end{equation}
where the factor $\sqrt{R_1/R_2}$ comes from using the power normalisation convention $|S_{11}|^2+|S_{21}|^2 = 1$. Substituting the formulas for $V_1$, $V_2$ and $S_{11}$ yields
\begin{equation}\label{eq:ABCD-to-S21}
    S_{21}(f) = \frac{2\sqrt{R_1/R_2}}{A+B/R_2+CR_1+(R_1/R_2)D},
\end{equation}
which results in the same formula as in Ref. \cite{Frickey1994}.
Furthermore, the comparison between Eq. \ref{eq:ABCD-to-H} and Eq. \ref{eq:ABCD-to-S21} shows that
\begin{equation}
    H(f) = \frac{1}{2}\sqrt{R_2/R_1}S_{21}(f).
    \label{eq:H-to-S}
\end{equation}

An important special case of Eq. \ref{eq:ABCD-to-S21} occurs when the circuit inside the `black-box' consists only of series components. Then the ABCD matrix is given by
\begin{equation}
\begin{pmatrix}
A & B
\\C & D
\end{pmatrix}
=
\begin{pmatrix}
1 & Z_B
\\0 & 1
\end{pmatrix},
\end{equation}
where $Z_B$ is the total series impedance of the black-box. Substituting this form into Eq. \ref{eq:ABCD-to-S21} gives
\begin{equation}
    \label{eq:series}
   \lvert S_{21}(f) \rvert^2 = \frac{4R_1 R_2}{\lvert R_1+R_2+Z_B \rvert^2},
\end{equation}
which agrees with the formula derived in Ref. \cite{Pascal2011} through a different method.

Similarly, we can consider a black-box in which all the components are connected in parallel. In this case
\begin{equation}
\begin{pmatrix}
A & B
\\C & D
\end{pmatrix}
=
\begin{pmatrix}
1 & 0
\\1/Z_B & 1
\end{pmatrix},
\end{equation}
where $1/Z_B$ is total admittance of the parallel elements of the black-box. Again, the substitution into Eq. \ref{eq:ABCD-to-S21} yields a useful simplification

\begin{equation}
\label{eq:parallel}
   \lvert S_{21}(f) \rvert^2 = \frac{4(1/R_1)(1/R_2)}{\lvert 1/R_1 + 1/R_2 + 1/Z_B \rvert^2},
\end{equation}
which has been derived and used in Ref. \cite{Thomas2019} to study the heat transport across a Josephson junction.

Importantly, Eq. \ref{eq:series} and Eq. \ref{eq:parallel} can also be applied in the case of complex terminating impedances, if the reactive/susceptive components are included into the black-box. Writing the complex forms explicitly gives
\begin{equation}
    \label{eq:seriesZ}
    \lvert S_{21}(f) \rvert^2 = \frac{4\mathrm{Re}[Z_1]
    \mathrm{Re}[Z_2]}{\lvert Z_1+Z_2+Z_B \rvert^2}
\end{equation}
and
\begin{equation}
\label{eq:parallelZ}
   \lvert S_{21}(f) \rvert^2 = \frac{4\mathrm{Re}[1/Z_1]\mathrm{Re}[1/Z_2]}{\lvert 1/Z_1 + 1/Z_2 + 1/Z_B \rvert^2},
\end{equation}
where $Z_1$ and $Z_2$ are the complex terminating impedances.

\section{Ports in Sonnet}\label{app:Port}

\begin{figure*}[h!]
\centering
\includegraphics{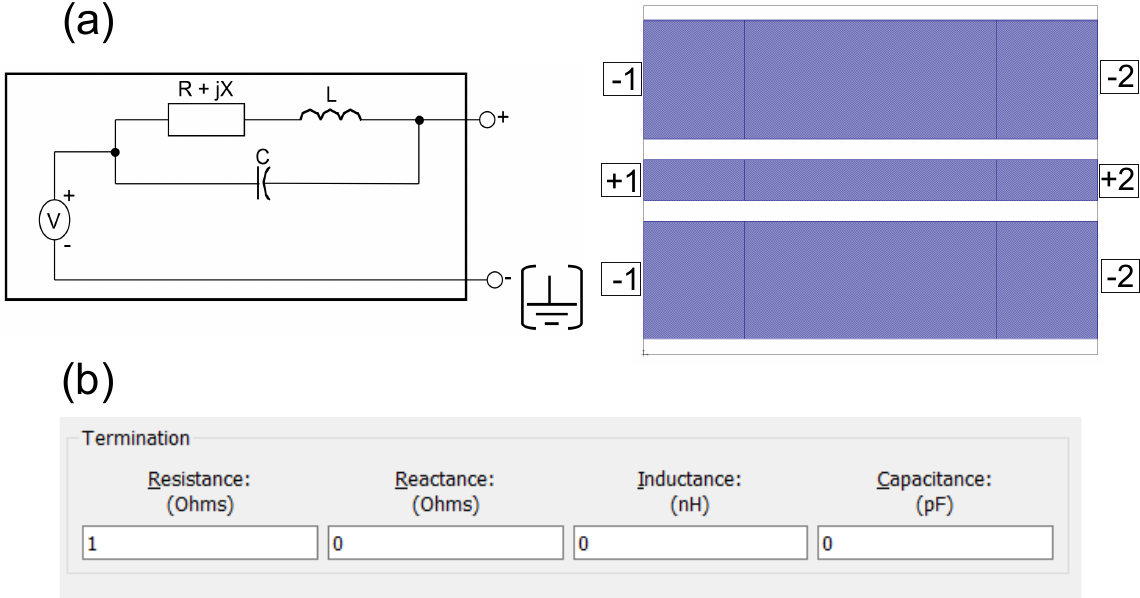}
\caption{Snapshot of the port configuration in Sonnet. The port termination can be set to in any combination of resistive and reactive elements. For our heat transport study, we only set and vary the resistive value $R$, that terminates the circuit (+) to the ground (-).\label{fig:A2}}
\end{figure*}

The port structure in Sonnnet consists of a voltage source in series with a normalising impedance component as shown in Fig. \ref{fig:A2}(a). By default, the port impedance has only a resistive component $R = 50~\Omega$. The setting can be overwritten by the user, and in our simulations we change and vary the resistive component $R$ while keeping the other component values at zero (Fig. \ref{fig:A2}(b)). Additionally to the resistor $R$, here we can also set a value of shunting capacitor $C$, series reactance $X$ and series inductor $L$. This option is important in the situation when the dimension of the resistor is significant and it cannot be assumed as a lumped element anymore, and the resistor's geometry starts to affect the wave propagation across it.

\section{Energy Spectrum for Spectroscopy Fitting}

\subsection{Hamiltonian of QHV Circuit}

The transmon qubit Hamiltonian
\begin{equation}\label{eq:Hamiltonian}
    H_Q = 4E_C(\hat{n} - n_g)^2 - E_\textrm{J}(\Phi)\cos{\hat{\phi}},
\end{equation} 
where $\hat{n}$ and $\hat{\phi}$ are the charge number and phase operator respectively. The parameter $n_g$ is the gate offset-charge.

The total Hamiltonian of a transmon coupled to two resonators with equal frequencies $\omega_1=\omega_2$,
\begin{equation}\label{eq:HamiltonianQHV}
    H =  H_Q + \sum\limits_{i = 1,2}(H_{R,i} + H_{I,i}) + \tilde{g}_{12} (a_1^\dag a_2 + a_2^\dag a_1),
\end{equation} 
where the Hamiltonian of each resonator, for $i\in\{1,2\}$, is
\begin{equation}\label{eq:Hamiltonian}
  H_{R,i}=\hbar \omega_i a_i^\dag a_i
\end{equation} 
and for the resonator-qubit interaction
\begin{equation}\label{eq:Hamiltonian}
  H_{I,i}=g_i \hat{n} (a_ic +a_i),
\end{equation}
with $a_i^\dag$, $a_i$ denoting the creation and annihilation operators. The parameters $\tilde{g}_{12}$ and $g_i$ denote the resonator cross-coupling and the coupling between qubit and resonator $i$, respectively.

\subsection{Hamiltonian of Double Pole QHV Circuit}

The two transmon Hamiltonians are, for $i\in\{\alpha,\beta\}$,
\begin{equation}\label{eq:Hamiltonian}
    H_{Q,i} = 4E_{C,i}(\hat{n}_i - n_{g_{i}})^2 - E_{\textrm{J,i}}(\Phi)\cos{\hat{\phi_{i}}}, 
\end{equation} 
where both transmons are identical. Total Hamiltonian of two transmons coupled to two identical resonators
\begin{equation}\label{eq:Hamiltonian2QHV}
    H =  \sum\limits_{i = \alpha,\beta} H_{Q,i} + \sum\limits_{i = 1,2}(H_{R,i} + H_{I,i}) + \tilde{g}_{12} (a_1^\dag a_2 + a_2^\dag a_1) + H_{Q,\alpha \beta},
\end{equation} 
where the Hamiltonian of each resonator, for $i\in\{1,2\}$, is
\begin{equation}\label{eq:Hamiltonian}
  H_{R,i}=\hbar \omega_i a_i^\dag a_i
\end{equation} 
and for resonator-qubit interaction
\begin{equation}\label{eq:Hamiltonian}
  H_{I,i}=g_i \hat{n} (a_i^\dag +a_i)
\end{equation} 
Qubit-qubit interaction
\begin{equation}\label{eq:Hamiltonian}
  H_{Q,\alpha \beta}=\tilde{g}_{\alpha \beta} (\hat{n}_a^\dag \hat{n}_b + \hat{n}_b^\dag \hat{n}_a)
\end{equation}Here the transmon-1 to resonator-2, transmon-2 to resonator-1, and resonator-1 to resonator-2 interactions are taken to be negligible. 

\section{Sonnet Simulation and ABCD model Parameters}

%The Sonnet simulation parameters used in section 4.
%$l_1=4350~\mathrm{\mu m}$ and $l_2=4350~\mathrm{\mu m}$,
%frequency 6.2
%$20~\mathrm{\mu m}$ centre conductor, with $10~\mathrm{\mu m}$ gaps to the ground plane. 
%$500~\mathrm{\mu m}$ vacuum layer
% from $0~\,\mathrm{GHz}$ to $10\,\mathrm{GHz}$
%The output capacitor is simulated using COMSOL to be $C_r = 23~\mathrm{fF}$.
%We estimated the distributed elements to be $C_l=171~\mathrm{pF/m}, L_l=405~\mathrm{nH/m}, \text{and}~Z_0=49~\mathrm{\Omega}$.
%and thus the propagation constant $\gamma$ only has phase constant $\beta =\omega l \sqrt{C_l L_l}$.
%

\begin{table}[h!]
\centering
\begin{tabular}{||c c||} 
 \hline
 Parameter & Value\\ [0.5ex] 
 \hline\hline
 Inductance per unit length, $L_l$ & $405~\mathrm{nH/m}$ \\ 
 Capacitance per unit length, $C_l$ & $171~\mathrm{pF/m}$  \\
 $l_1$  & $4723~\mathrm{\mu m}$  \\
 $l_2$  & $580~\mathrm{\mu m}$  \\
 $C_r$ &  $23~\mathrm{fF}$ \\[1ex] 

 \hline
\end{tabular}
\caption{Simulation parameters for curves at Fig~\ref{fig:2}.}
\label{table:1}
\end{table}

\begin{table}[H]
\centering
\begin{tabular}{||c c||} 
 \hline
 Parameter & Value\\ [0.5ex] 
 \hline\hline
 Inductance per unit length, $L_l$ & $405~\mathrm{nH/m}$ \\ 
 Capacitance per unit length, $C_l$ & $171~\mathrm{pF/m}$  \\
 $l$  & $5119~\mathrm{\mu m}$  \\
 $C_r$ &  $10~\mathrm{fF}$ \\ 
 $C_s$ &  $96~\mathrm{fF}$ \\
 $I_{C\Sigma}$ &  $72~\mathrm{nA}$ \\ [1ex] 
 \hline
\end{tabular}
\caption{Simulation parameters for curves at Fig~\ref{fig:3}.}
\label{table:1}
\end{table}

\begin{table}[h!]
\centering
\begin{tabular}{||c c||} 
 \hline
 Parameter & Value\\ [0.5ex] 
 \hline\hline
 Inductance per unit length, $L_l$ & $405~\mathrm{nH/m}$ \\ 
 Capacitance per unit length, $C_l$ & $171~\mathrm{pF/m}$  \\
 $l$  & $5119~\mathrm{\mu m}$  \\
 $C_r$ &  $10~\mathrm{fF}$ \\ 
  $C_t$ &  $20~\mathrm{fF}$ \\
   $C_s$ &  $61~\mathrm{fF}$ \\
    $I_{C\Sigma}$ &  $72~\mathrm{nA}$ \\ [1ex] 
 \hline
\end{tabular}
\caption{Simulation parameters for curves at Fig~\ref{fig:5}.}
\label{table:1}
\end{table}

\section*{References}
\bibliographystyle{iopart-num}

\bibliography{ReferencesSONNET_HT.bib}

\end{document}